\documentclass[11pt, a4paper]{article}
\usepackage{amsfonts}
\usepackage{amsmath}
\usepackage{breqn}
\usepackage{bm}
\usepackage{graphicx} 
\usepackage{lscape}
\usepackage{hyperref}
\usepackage{amssymb}
\usepackage{amsthm}
\usepackage{rotating}
\usepackage{color}
\usepackage[table]{xcolor}
\usepackage{endnotes}
\usepackage{booktabs}
\usepackage{threeparttable}
\usepackage{longtable}
\usepackage{float}
\usepackage[square]{natbib}

\definecolor{LightGray}{gray}{0.95}
\definecolor{LightBlue}{rgb}{0.85,0.93,1.00}
\providecommand{\U}[1]{\protect\rule{.1in}{.1in}} \textwidth=420pt
\RequirePackage{amsopn} \RequirePackage{amsfonts}
\RequirePackage{amsthm}

\newtheorem{Estimation assumption}{Estimation assumption}

\begin{document}

\title{Evolution of global inequality in well-being: \\
A copula-based approach}
\author{Koen Decancq\thanks{%
Herman Deleeck Centre for Social Policy, University of Antwerp} \and Vanesa
Jord\'a \thanks{%
Department of Economics, University of Cantabria}}
\date{}
\maketitle

\begin{abstract}
We employ a flexible parametric model to estimate global income, health, and education distributions from 1980 to 2015. Using these marginal distributions within a copula-based framework, we construct a global joint distribution of well-being. This approach allows us to specifically analyze the impact of dependency structures on global well-being inequality. While inequality decreased in each individual dimension, our findings suggest that multidimensional inequality does not necessarily follow this trend. Its evolution is influenced by the interdependence among dimensions and the chosen inequality aversion parameter.

\end{abstract}

\section{Introduction}
The analysis of economic inequality, traditionally focused on purely monetary outcomes, has a long history in economics. While there is a growing body of work on the evolution of global income inequality, much existing scholarship has been hampered by the scarcity of individual-level data. To address these limitations, most empirical studies on global inequality rely on grouped data, typically available as income shares across numerous countries \citep{bourguignon2002, lakner2016, nino2014, dowrick2005, anand2016}. Consequently, these analyses often depend on a limited number of data points, fueling significant criticism that past estimates of global inequality may suffer from severe biases. In such cases, parametric models can provide more accurate estimates even with minimal data \citep{jorda2019, jorda2020}.

Another key limitation in the existing research on global inequality is its predominantly income-centered focus. Many scholars have challenged the primacy of economic metrics in measuring well-being, arguing that well-being encompasses other equally relevant dimensions that are not adequately captured by economic data alone \citep{alkire2002, stiglitz2009}. This multidimensional view of well-being presents an even greater challenge for estimating global inequality, as data on non-economic variables are often available only as national aggregates that only provide information on the average achievement in a particular dimension.

Due to these data constraints, most previous efforts to assess global well-being inequality have relied on country-level data to compute population-weighted inequality measures (e.g., \citealp{decancq2011, mcgillivray2004}). Although this approach is straightforward, it risks overlooking substantial disparities within countries, as individuals differ widely in earnings, educational attainment, and health status. While cross-country differences in per capita income largely account for global income inequality, prior research indicates that within-country disparities contribute over 90 percent to global inequality in non-monetary dimensions such as health and education \citep{smits2009, jorda2017}. Thus, neglecting intra-country differences in well-being could result in highly biased estimates.

To address the challenge of measuring within-country variation in well-being, previous studies have generally used data grouped by income \citep{grimm2008} or at sub-national levels \citep{kummu2018, smits2019}. The study by \citet{harttgen2012} is unique in estimating inequality in human development at the household level, employing a survey-based approach to assess household well-being. Given the limited data needed to construct dimension-specific indices, they use parametric models to estimate the three components of the Human Development Index (HDI). Despite the advantages of parametric models in data-constrained settings, the approach proposed by \citet{harttgen2012} could only be applied to a select number of countries, as surveys rarely capture income, education, and health data concurrently. This paper proposes a more general approach, also based on parametric models, which enables the estimation of the global distribution of well-being. Drawing on grouped data on income, health, and education, we employ a copula-based model to estimate multidimensional well-being inequality from 1990 to 2010.

The rest of the paper is organized as follows. Section 2 presents our empirical strategy, including a discussion of the data and methods used to estimate national and global distributions of well-being. Section 3 examines multidimensional dependence measures and the multidimensional inequality index used to quantify global well-being inequality. Section 4 details the results of our analysis, while Section 5 concludes with a discussion of the implications of our findings.

\section{Modeling the joint well-being distribution}
The notion of well-being as a multidimensional construct that encompasses aspects beyond economic growth is well-established in the economic literature. Consequently, scholars have become increasingly focused on exploring ways to synthesize different aspects of well-being into a composite index. Among the potential candidates, the HDI stands out as the most prominent indicator, capturing well-being not only through income capacity but also by including education and health opportunities. Its popularity lies in its simplicity and broad geographical coverage, which allows for comparisons across countries in terms of well-being over an extended period.

Since the 1990s, nearly all countries have achieved substantial improvements in human development, with people generally becoming better educated and enjoying higher standards of living over longer lifespans. Yet, these averages may conceal imbalances that benefit privileged groups at the expense of others. To provide a complete picture of the distribution of well-being, it is essential to incorporate the multiple dimensions of well-being explicitly into inequality analysis. Using the HDI as a theoretical framework, the joint density function of income ($Y$), health ($H$), and education ($X$) can be defined as follows \citep{sklar1959}:

\begin{equation}\label{sklar} f(y,h,x) = f_Y(y) \ast f_H(h) \ast f_X(x) \ast c\left(F_Y(y),F_H(h),F_X(x)\right) \end{equation}

where $f_Y(y)$, $f_H(h)$, and $f_X(x)$ are the probability density functions (pdf) of income, health, and education, respectively. Thus, the multivariate distribution of well-being depends on the marginal distributions of each dimension, represented by the product of their pdfs, and the dependence structure between dimensions, represented by the pdf of the copula function evaluated at the marginal cumulative distribution functions (cdf), denoted by $F_j(.)$, where $j = Y, H, X$. Therefore, a copula is a multivariate distribution with uniform marginals supported on $[0,1]^n$.

Previous research on well-being inequality has often focused on analyzing the distribution of each dimension independently \citep[see, e.g.,][]{mcgillivray2010, ram1992}. This approach provides only a limited perspective, overlooking a critical source of inequality: the interrelations between dimensions. If we aim to incorporate multiple dimensions into the analysis of inequality, it is necessary to also account for the dependencies between them.

In this paper, we employ a copula-based approach to estimate Eq. (\ref{sklar}). Our model leverages national and regional data to estimate the copula parameters. As for the marginal distributions, these would be relatively straightforward to estimate if individual or household data were available. Unfortunately, we face significant data limitations for the three HDI dimensions. Nonetheless, certain summary statistics on the distribution of income, health, and education are periodically released and provide some information.

When only grouped data are available, parametric models tend to yield more reliable estimates than alternative methods \citep[see][]{dhongde2013, jorda2017}. Therefore, despite the widespread use of nonparametric techniques in inequality analysis, we opt for parametric methods to estimate the marginal distributions. We draw upon the existing literature on parametric models to represent the three HDI dimensions, acknowledging that each may follow distinct distributional patterns. Accordingly, we specify different functional forms for each dimension to capture the unique characteristics of each variable and thereby provide an accurate picture of the distributional patterns of the three indicators.\footnote{Although the same functional form is used to model each dimension across all countries, we have selected highly flexible models to minimize misspecification bias. Our model captures country-specific patterns and their evolution over time, as parameter estimates vary by country and year.}

\subsection{The size distribution of income}
A wide array of parametric distribution families has been proposed in the literature to model income distributions.\footnote{For a comprehensive review, refer to Kleiber and Kotz (\citeyear{kleiber2003}).} Among these, the generalized beta distribution of the second kind (GB2) is considered one of the most promising options \citep{jenkins09, mcdonald1995, mcdonald1995m}. The GB2 distribution is defined by the probability density function (pdf) ($a,b,p,q \geq 0$) as follows \citep{Mcdonald1984}:

\begin{equation} f(y;a,b,p,q)=\frac{ay^{ap-1}}{b^{ap}B(p,q)[1+(y/b)^{a}]^{p+q}},;;y\geq 0, \label{Dgb2} \end{equation}%

where $B(p,q)=\int_{0}^{1}t^{p-1}(1-t)^{q-1}dt$ is the beta function. Parameters $a$, $p$, and $q$ are shape parameters, and $b$ is a scale parameter.

The GB2 distribution is a general model that encompasses many parametric forms commonly used in income inequality research. Due to its flexibility, this model can provide a robust approximation of income patterns and yield reliable estimates of inequality measures, even when only a few data points are available \citep{jorda2020}. For example, distributions with a mode at zero are typical of developing countries with high poverty rates, while as the middle class emerges, distributions often shift to a single-mode pattern. Although we use the same functional form across countries, different distributional patterns can be captured since parameter estimates vary by country and over time.\footnote{\citet{Shorrocks08}'s approach is a method to generate microdata samples with the same quantiles used for estimation. We do not employ this method here, as our estimation criterion is to minimize deviations from observed income shares. Given that we simulate samples of size 10,000, this approach would not significantly alter the results.}

In defining our estimation strategy, understanding the data structure is essential. Let \textbf{y} be an \textit{i.i.d.} random sample of size $N$ from a continuous income distribution over the support $T = [0, \infty)$. Often, however, data is not available at the individual level for most countries, but grouped data on income shares, such as five or ten points from the Lorenz curve, are publicly accessible for a large sample of countries. Suppose $T$ is divided into $J$ mutually exclusive intervals, $T_j = [t_{j-1}, t_j), j=1, \dots, J$, and denote by $s_j = \sum_{i=1}^N \bm{1}{[0, t_j)}(y_i)y_i/\sum{i=1}^N y_i, j=1, \dots, J$ the income share held by the population proportion $u_j = \sum_{i=1}^N \bm{1}_{[0, t_j)}(y_i)/N, j=1, \dots, J$. When income shares are provided, the shape parameters of the distribution ($a, p, q$) can be estimated by minimizing the squared deviations between observed and theoretical points on the Lorenz curve for the GB2 distribution:\footnote{We use the \texttt{GB2group} R package to estimate the GB2 distribution with an equally weighted minimum distance estimator, given by Eqs. (\ref{NLSincome}) and (\ref{media}). For details on the optimization algorithm, refer to \citet{jorda2020}.}

\begin{equation} \label{NLSincome} \min_{a,p,q} \sum_{j=1}^{J-1}\left[B\left(B^{-1}(u_j;p,q);p+\frac{1}{a},q-\frac{1}{a}\right) -s_j\right]^2, \end{equation}

where $q > 1/a$ and $B^{-1}(x;p,q)$ is the inverse of the incomplete beta function ratio, given by $B(v;p,q)=\int_0^v t^{p-1}(1-t)^{q-1} , dt/B(p,q)$.

Eq. (\ref{NLSincome}) does not involve the scale parameter $b$, as the Lorenz curve is scale-invariant. To estimate the scale parameter, we propose solving the following equation:

\begin{equation} \label{media} \bar{Y} = b \frac{B(\hat{p} + \frac{1}{\hat{a}}, \hat{q} - \frac{1}{\hat{a}})}{B(\hat{p}, \hat{q})}, \end{equation}%

which aligns per capita income ($\bar{Y}$) with the mean of the GB2 distribution, evaluated at the parameter estimates from Eq. (\ref{NLSincome}).

Once we estimate each country’s income distribution using Eqs. (\ref{NLSincome}) and (\ref{media}), we construct the global income distribution as a mixture of national distributions weighted by population:

\begin{equation} \label{cdf_income} F(y; \mathbf{a}, \mathbf{b}, \mathbf{p}, \mathbf{q}, \boldsymbol\eta) = \sum_{c=1}^C \eta_c B\left(\frac{(y/b_c)^{a_c}}{1+(y/b_c)^{a_c}};p_c,q_c\right), \end{equation}

where $\eta_c$ is the population weight of country $c$, for $c=1, \dots, C$.

Thus, to estimate the global distribution of income, we require per capita income data and grouped income shares. The choice between mean incomes from national accounts or household surveys is a key decision in global inequality analysis. While prior research has mostly relied on national account data due to limited survey data on means (e.g., \citealp{atkinson2010, bhalla2002, bourguignon2002, dowrick2005}), some exceptions exist (e.g., \citealp{anand2008, milanovic2011, lakner2016}). We approximate mean income by per capita Gross National Income (GNI), adjusted for purchasing power parity (PPP) at constant 2011 prices, from the World Bank's World Development Indicators.\footnote{http://data.worldbank.org/indicator/NY.GNP.PCAP.PP.KD} Since this indicator estimates residents' income, it is linked to income ownership. In contrast, Gross Domestic Product (GDP) measures income within a country’s borders, relating to geographic location.

Data on income shares are gathered from the World Income Inequality Database (WIID) version 3.4 (UNU-WIDER, \citeyear{WIID2014}), the most reliable and comprehensive distributional dataset. WIID provides cross-country information on Gini indices and income (or consumption) shares for 182 countries from 1867 to 2015. Our analysis focuses on 1990-2010 in ten-year intervals (1990, 2000, 2010). When data for the exact year is unavailable, we include observations within four years of each target year. To minimize potential issues arising from conceptual differences in WIID data regarding the unit of analysis, welfare concept, and population coverage, we select comparable observations within countries. Since our analysis centers on global interpersonal inequality, we prioritize nationally representative survey data, considered the highest quality in WIID. For the welfare concept, we prefer income-based over consumption-based data to align with national account mean incomes. Finally, we use per capita household income data rather than adult-equivalent adjustments.

\subsection{The size distribution of educational outcomes}

Let $X$ be a continuous variable representing the years of schooling completed until either the maximum level of education is achieved or schooling is discontinued. Because microdata on the exact years of schooling completed by individuals is rarely available, it is common to rely on grouped data to estimate the distribution of education. Individuals are typically categorized into four broad educational levels according to their attainment: no schooling (NS), primary (P), secondary (S), and tertiary (T) education. These levels are further classified into completed (C) and incomplete (I) groups, depending on whether the educational cycle was finished, to provide more disaggregated data. Let $r_j$, where $j = \text{NS, PI, PC, SI, SC, TI, TC}$, denote the proportion of the population that achieved each educational level $j$. This information is combined with the official duration of each educational level in a given country ($d_j$) to transform the categorical variable into a continuous indicator, following methods in studies such as \citet{barro2013}, \citet{cohen07}, and \citet{morrisson09}.

UNESCO’s Statistical Yearbook provides country-specific figures for the duration of each educational level. However, for incomplete levels, there is no data on the exact years individuals attended, so an arbitrary value is assigned for all individuals in this category. This subjective choice may introduce some measurement error. The parametric approach developed by \citet{jorda2017} avoids these discretionary assumptions and provides an accurate representation of the distribution of education. They propose using the generalized gamma (GG) distribution to model the distribution of schooling years, defined by the pdf as follows \citep{Stacy1962}:

\begin{equation}\label{GGcox} f(x;a,b,p)=\frac{ax^{ap-1}e^{-(x/b)^a}}{b^{ap}\Gamma(a)},;;x\ge 0, \end{equation}%

where $\Gamma(a)=\int_0^\infty x^{a-1}e^{-x} , dx$ is the gamma function, $b>0$ is a scale parameter, and $a>0$ and $p>0$ are shape parameters. This parametric model can accommodate both one-mode and zero-mode distributions. One mode is common in developed countries with compulsory schooling, while zero-mode distributions are typical in developing countries with high illiteracy rates.

We use a minimum distance estimator to estimate the education distribution in each country. Attainment rates $(r_j)$ represent the empirical cumulative probabilities of achieving each level of education, i.e., attending $d_j$ years of schooling or less. The strategy is to minimize the sum of squared deviations between attainment rates and the theoretical probabilities for each educational level $(Pr[X \leq d_{j}])$, modeled by the cumulative distribution function (cdf) of the GG distribution. We also account for the censored nature of attainment rate data. Although individuals classified as having completed tertiary education are assumed to have completed university, some may have enrolled in master's or Ph.D. programs. To address this right censoring, we model the last category using the survival function $(1-F(x))$ rather than the cdf used for the lower levels. Thus, for each country and year, the function to be minimized is:

\begin{equation}\label{minRSS} \min_{a,b,p}\sum_{j=1}^{J-1}\left[IG\left((d_{j})^a;p,b^a\right)-r_{j}\right]^2+ \left[1-IG((d_{TC})^a; p,b^a)-r_{TC}\right]^2, \end{equation}%

where $IG (.)$ denotes the incomplete gamma function.

The global distribution of education for each year is constructed from national estimates, obtained using Eq. (\ref{minRSS}), as a mixture of distributions:

\begin{equation}\label{cdf_educ} F(x; \mathbf{a}; \mathbf{b}; \mathbf{p}; \boldsymbol\eta)= \sum_{c=1}^{C}\eta_cIG\left[x^{a_c};p_c,(b_c)^{a_c}\right], \end{equation}

where $\eta_c$ is the population weight of country $c$.

Data on the official years of schooling for primary and secondary levels are taken from UNESCO \citep{UNESCO}. Following UNESCO guidelines, the illiterate population is assumed to have attended less than one year of schooling. For tertiary education, we assume a standard duration of four years across all countries and periods.\footnote{Most previous studies also adopt this assumption \citep{cohen07, barro2013} due to significant variation in the duration of tertiary education programs. On average, completed short-cycle tertiary education (ISCED 5) took 3.7 years between 2000 and 2010 \citep{UNESCO2013}.}

We use the Barro-Lee data on educational attainment rates for the total population aged 15 and over \citep{barro2013}. These figures cannot directly estimate $r_j = Pr[X \leq d_j]$ as they represent $Pr[X \leq d_j | A \geq 15]$, where $A$ is a continuous variable for age. Since demographic data is generally broken into five-year intervals, we estimate attainment rates for three age groups: 0-4, 5-9, and 10-14. To approximate education distribution for these age groups, we make the following assumptions:

The population aged 0-4 is assumed to have no schooling.
The enrollment ratio for primary education serves as a proxy for the proportion of the population aged 5-9 classified as incomplete primary. The remainder is assumed to be illiterate.
For the 10-14 age group, stronger assumptions are needed. For example, if schooling starts at age 6 and primary education lasts six years, the 10-11 age group can only be classified as primary or illiterate, while ages 12-14 may also include those who started secondary education. Two subgroups are identified: those who could only have attended or completed primary education, classified using the primary enrollment ratio, and those who may have begun secondary education, classified using the secondary enrollment ratio. Ages within this interval are assumed to be uniformly distributed, meaning all ages are equally likely.
To obtain the unconditional distribution of years of schooling from conditional probabilities, we estimate the attainment rate for each level of education as follows:


$$r_j=Pr[X\leq d_j]=\sum_{k=1}^{K}Pr[E \leq d_j| A\in a_k]Pr[A\in a_k], $$
where $a_k, k = 1, 2, 3$ are the four groups of age: from 0 to 4, from 5 to 9, from 10 to 14 and over-15 and $Pr[A\in a_k]$ is estimated using data from World Population Prospects (UN, \citeyear{desa2017}). 

\subsection{The distribution of life spans}



Before discussing the measurement of the distribution of length of life, it is important to note that, although the data is available in an aggregated form, there are 100 data points for each country and year. This granularity allows for a non-parametric approach, which can provide an accurate approximation of the distribution of length of life. 

Let $H$ be a random variable representing the lifespan of an individual until death. Consider $J$ realizations of that variable, $h_1, \dots, h_J$, grouped in $K$ age intervals, $[a_0, a_1),$ $\dots, [a_{K-1}, a_K)$. Period life tables provide information on the survivals at the beginning of each interval $(l_{k})$ of a hypothetical birth cohort of a synthetic population of $100,000$ individuals. The number of individuals who died aged between $a_{k-1}$ and $a_k$ can be obtained straightforwardly as, $d_k=l_{k-1}-l_k$. Then, dividing these figures by the size of this hypothetical population we obtain the probability of dying at each interval,
$$f_k = \frac{d_k}{100,000}.$$
Hence, $f_k$ are points of the probability density function (pdf) of the distribution of length of life, which gives the probability of dying aged between $x_{k-1}$ and $x_k$. Since no information on the distribution of deaths within the age intervals is available, we assume that the population dies uniformly throughout each interval.\footnote{Despite being a conventional assumption, this approach would underestimate inequality levels because differences between individuals that belong to the same age intervals are not considered. This potential limitation might not affect our results substantially because the data is provided for 100 intervals.}

We can compute the pdf of the distribution of length of life for any group of countries as a mixture of the national distributions, weighted by their population shares. Let $H^{(i)}$ be the length of life in the country $i, i=1,...,N$, and $f_k^{(i)}$ be the mortality rate of the age group $k$ in the country $i$. Then, the global mortality rate is given by:
\begin{equation}\label{mixture}
f_k^{(R)}=\sum^{C}_{c=1}\eta_cf_k^{(c)}, k =1, \dots, K,
\end{equation}
where $\eta_c$ stands for the population weight of the country $c, c=1,...,C$.


Period life tables were retrieved from the World Population Prospects: The 2021 Revision, developed by the UN Population Division \citep{desa2017}. This database is particularly valuable due to its extensive geographical and temporal coverage, providing a balanced panel of period life tables for 201 countries/areas from 1950 to 2010, in five-year intervals. Given the frequency of this data, our benchmark years--1980, 1985, 1990, 1995, 2000, 2010, and 2015--overlap with two different life tables. For example, the period 1985-1990 spans from July 1, 1985, to June 30, 1990, while the next table covers July 1, 1990, to June 30, 1995. To improve comparability between health and income data, we select the life table period that includes the actual year of the income observation. In cases where income data exactly corresponds to one of the benchmark years, we use the table from the most recent period.

 \subsection{The copula: A model for the dependence structure}\label{copula}

Thus far, we have introduced the parametric models used to represent the marginal distributions in Eq. (\ref{sklar}). We now shift focus to estimating the dependence structure between the three dimensions of well-being. Regardless of the specific forms of the marginal distributions, the dependence between income, health, and education is fully captured by the copula. A potential limitation of the copula-based approach is that, theoretically, there are infinitely many possible copulas. While a growing body of research explores the use of copulas to measure multidimensional poverty and inequality \citep{decancq2014, tkach2018, perez2016}, there are virtually no published studies identifying parametric models that reliably represent the multivariate distribution of well-being. Selecting an appropriate copula generally involves considering both model features and data constraints, such as model tractability, the number of copula parameters, and the feasibility of obtaining consistent estimates with the available data.

Since there is no available information on the dependence structure between income, health, and education at the global level, we rely on two mild assumptions to establish inequality bounds for multivariate inequality:

\begin{enumerate} \item No dependence between the three dimensions. In this case, the joint distribution of well-being is characterized by an independent copula: \begin{equation} \Pi(y, x, h) = F(y) \cdot F(x) \cdot F(h) \end{equation}
This implies that there is no association between income, health, and education--knowing an individual's status in one dimension provides no information about their status in the others. This assumption represents the lowest possible level of multidimensional inequality, as it does not account for compounded advantages or disadvantages across dimensions.

\item Perfect positive dependence between the three dimensions of well-being. Here, the joint distribution is defined by:
\begin{equation}
M(y, x, h) = \min\{F(y), F(x), F(h)\}
\end{equation}
Under this assumption, individuals with high income also have high levels of health and education, and vice versa. This assumption represents the highest possible level of multidimensional inequality, as it maximizes the overlap between dimensions, assuming that all advantages or disadvantages are perfectly aligned.
\end{enumerate}

Given the marginal distributions, these two distributions yield the minimum and maximum bounds for multidimensional inequality. This approach, however, relies on the assumption that there is no negative dependence between dimensions of well-being. In reality, negative dependence—where higher attainment in one dimension could be associated with lower attainment in another—could theoretically occur. For example, individuals with higher education might, in some cases, experience lower income if they work in underpaid sectors, or those in physically demanding jobs may have high incomes but poorer health. While such cases may exist, we assume they do not significantly influence the global distribution to justify modeling negative dependence. Additionally, negative dependence structures could theoretically reduce overall multidimensional inequality by balancing inequalities across dimensions. 

Despite the lack of information about the dependence structure, evidence from a variety of sources provides support for a positive relationship between education and income (Becker and Chiswick, \citeyear{becker1966}; Miller, \citeyear{miller1960}). Prior research have also furnished empirical support for a socioeconomic gradient, along which citizens with higher income and education are better off in terms of health (Cutler et al., \citeyear{cutler2006}; Adams et al. \citeyear{adams2003}). Hence, without strong empirical evidence supporting substantial negative dependence across these dimensions at a global level, we focus on establishing bounds using assumptions of independence and perfect positive dependence.

To further examine the influence of dependence on the evolution of global well-being inequality, we consider intermediate cases between the two extremes presented above. Specifically, we use a mixture of the comonotonic and independent copulas:

\begin{equation}\label{mix_c} C_M(y, x, h) = (1 - \omega) , \Pi(F(x), F(y), F(h)) + \omega , M(F(x), F(y), F(h)), \end{equation}

where $\omega$ represents the weight assigned to the comonotonic copula. In this framework, $\omega$ serves as a dependence index, with $\omega \in [0, 1]$. Notably, it can be shown that the multivariate Spearman’s rank correlation for this copula is equal to $\omega$, further supporting its interpretation as a measure of dependence.

When $\omega = 0$, the copula reduces to the independent copula $\Pi(F(x), F(y), F(h))$, indicating complete independence between income, health, and education. In this case, well-being inequality is minimized because there are no reinforcing dependencies between dimensions. Conversely, when $\omega = 1$, the copula becomes the comonotonic copula $M(F(x), F(y), F(h))$, reflecting perfect positive dependence, where individuals with high values in one dimension are always high in the others. This results in maximum well-being inequality, as disparities in one dimension are fully mirrored across the others.

This approach offers a nuanced understanding of the role of dependence, as varying $\omega$ can reflect different societal structures or patterns of inequality where the relationships between income, health, and education are neither fully independent nor perfectly aligned. Adjusting $\omega$ will allow for sensitivity analysis, helping us assess how changes in dependence impact the overall level of well-being inequality.

\section{Dependence and multidimensional inequality}\label{dep_atk}

Let $W=\left(Y,H,X\right)$ be a random vector representing the three dimensions of well-being: income ($Y$), length of life ($H$), and educational outcomes ($X$). A realization of the random vector $\left(y,h,x\right)$ reflects an individual’s well-being outcomes. Let $F_{W}$ denote the joint distribution function of $W$ as given in (\ref{sklar}), with marginal distributions $F_{Y}$, $F_{H}$, and $F_{X}$ corresponding to the cdfs of the random variables. The averages of these random variables are denoted by $\mu_{Y}$, $\mu_{H}$, and $\mu_{X}$, respectively.

To measure inequality in the three-dimensional well-being distribution, we use a multidimensional generalization of the Atkinson inequality index:

\begin{equation}\label{mv_atk}
I(\bm{w})=1-\left[\frac{1}{n}\sum_{i=1}^n \left(\frac{U(y_i,h_i,x_i)}{U(\mu_{Y},\mu_{H},\mu_{X})}\right)^{1-\varepsilon}\right]^{1/(1-\varepsilon)},
\end{equation}
where 
\begin{equation}\label{gen_mean}
U(y,x, h)=\left[\omega_{Y}\times g_{Y}\left(y\right)^{1-\beta}+\omega_{H}\times g_{H}\left(h\right)^{1-\beta}+\omega_{X}\times g_{X}\left(x\right)^{1-\beta}\right]^{1/(1-\beta)}.
\end{equation}

This inequality index reflects a two-step procedure \citep{bosmans2015, decancq2017}. In the first step, equation (\ref{mv_atk}) is used to measure an individual’s well-being by aggregating their outcomes across the three well-being dimensions. This is done using a generalized mean, or constant elasticity of substitution (CES) aggregation formula [see \citep{decancq2013} for further discussion]. The parameters in this flexible aggregation formula reflect value judgments about the nature of well-being. 

In our empirical application, we use parameter values implicit in the HDI as a benchmark and also discuss the sensitivity of results to alternative parameter choices. The dimension-specific transformation functions $g_{Y}$, $g_{H}$, and $g_{X}$ convert each dimension’s outcomes to a unit-free index between 0 and 1. The weighting scheme $\left(\omega_{Y},\omega_{H},\omega_{X}\right)$ captures the relative importance of each dimension; in the HDI, these weights are equal, i.e., $\omega_{Y}=\omega_{H}=\omega_{X}=1/3$. The parameter $\beta$ indicates the degree of complementarity between the dimensions of well-being. Several special cases are defined by adjusting $\beta$. When $\beta=0$, the formula becomes additive, treating the dimension indices as perfect substitutes (as in the HDI prior to 2010). If $\beta$ approximates 1, the aggregation procedure becomes multiplicative, allowing a one percent decrease in one dimension to be offset by a one percent increase in another (as in the current HDI). When $\beta$ approaches infinity, the aggregation formula approaches the minimum operator.

In the second step, we measure multidimensional inequality using equation (\ref{gen_mean}). This aggregation formula is also based on a generalized mean, extending the one-dimensional Atkinson inequality index \citep{atkinson1970}. The parameter $\varepsilon$ captures the inequality aversion of this measure: higher values of $\varepsilon$ increase the weight placed on outcomes at the lower end of the distribution. In the limit, as $\varepsilon$ approaches infinity, the measure adopts a Rawlsian perspective, focusing solely on the outcomes of the worst-off individual.

The sensitivity of the multidimensional inequality index to dependence between well-being dimensions depends on the relationship between the inequality aversion parameter $\varepsilon$ and the degree of complementarity $\beta$. \citet{bourguignon1999} shows that when $\varepsilon>\beta$, an increase in dependence between dimensions results in higher multidimensional inequality. The higher the complementarity between dimensions (as set by $\beta$), the greater the inequality aversion ($\varepsilon$) must be for increased dependence to heighten inequality. When $\varepsilon=\beta$, the multidimensional inequality index becomes insensitive to dependence between dimensions, allowing multidimensional inequality to be computed solely from the marginal distributions. This approach, which has received attention in the literature \citep{foster2005, alkire2010}, has been used by the UNDP in defining the inequality-adjusted HDI. Although this is practically appealing, there is no \emph{a priori} reason why these two normative parameters should necessarily be equal, as they represent distinct aspects of multidimensional evaluation. For instance, an observer might adopt a Rawlsian perspective that focuses on outcomes for the least advantaged without needing to consider the dimension indices as perfect complements.

\section{Results}

In this section, we present the evolution of global well-being inequality under varying levels of dependence between the three dimensions: income, health, and education. Our sample comprises 78 countries, covering approximately 80 percent of the global population from 1990 to 2010. 

\subsection{Global unidimensional inequality: income, health and education}

\begin{figure}
\caption{The global distribution of income, health and education\label{Figure3}}
\begin{tabular}{cc}
 \includegraphics[scale = 0.1]{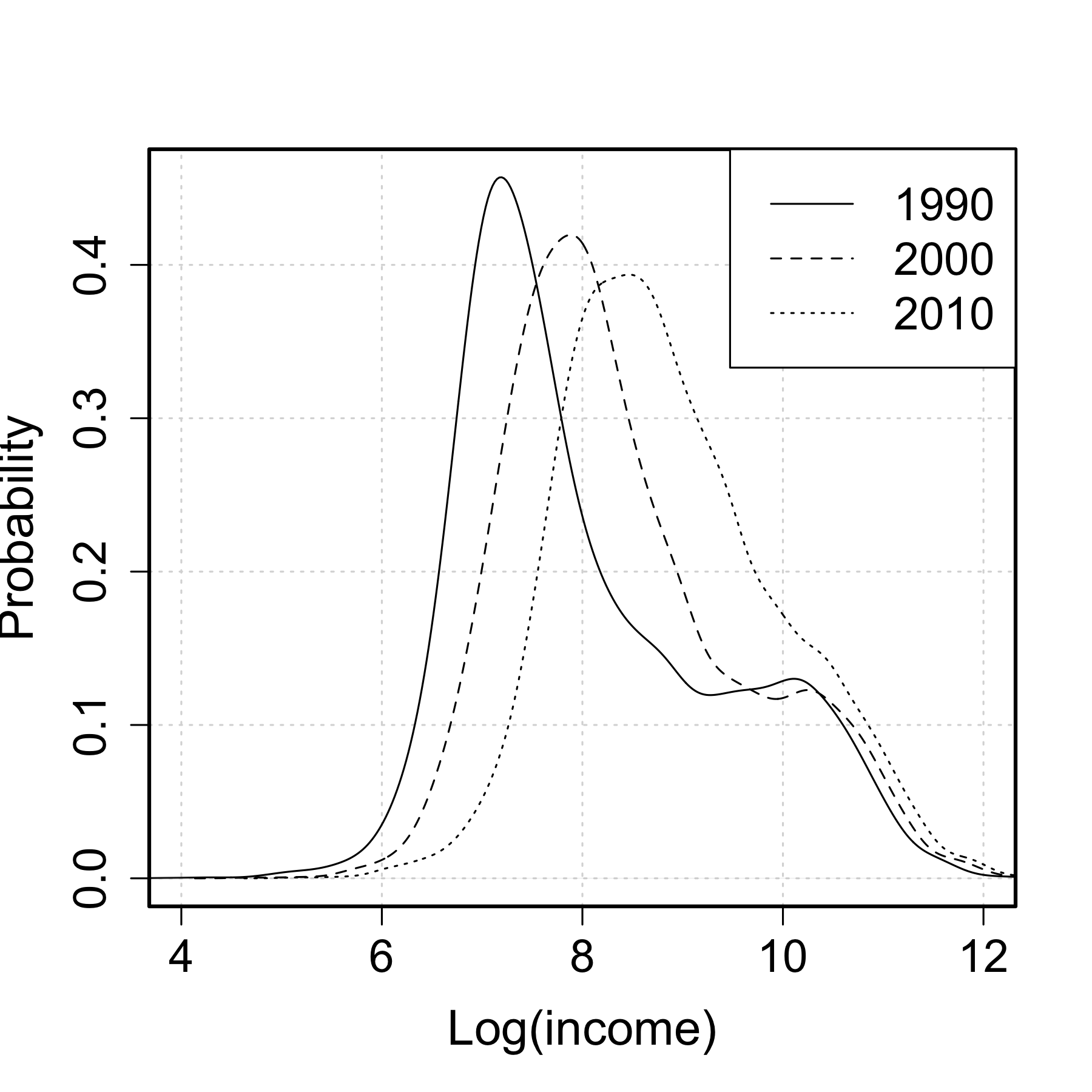}  & \includegraphics[scale = 0.1]{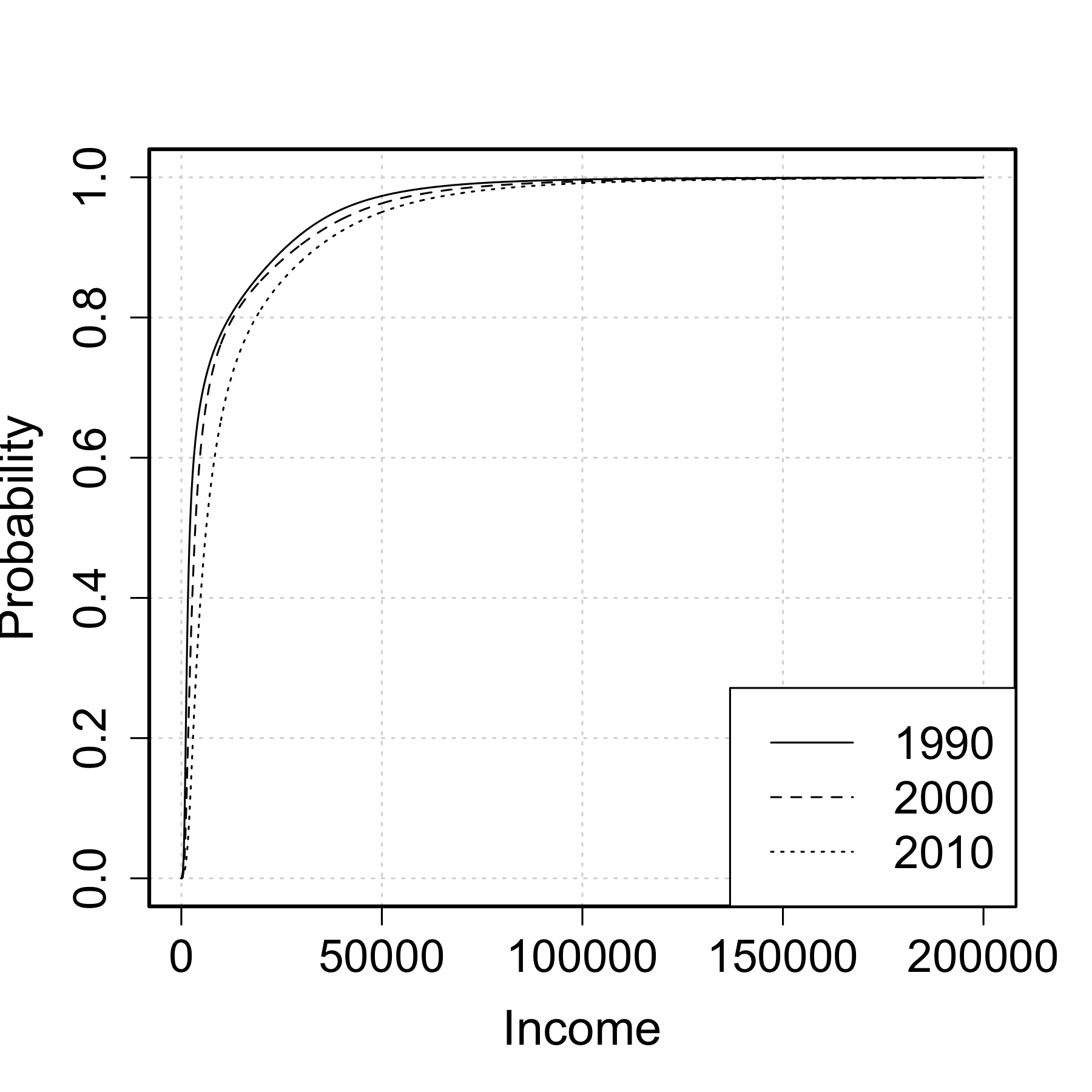}\\
  \includegraphics[scale = 0.1]{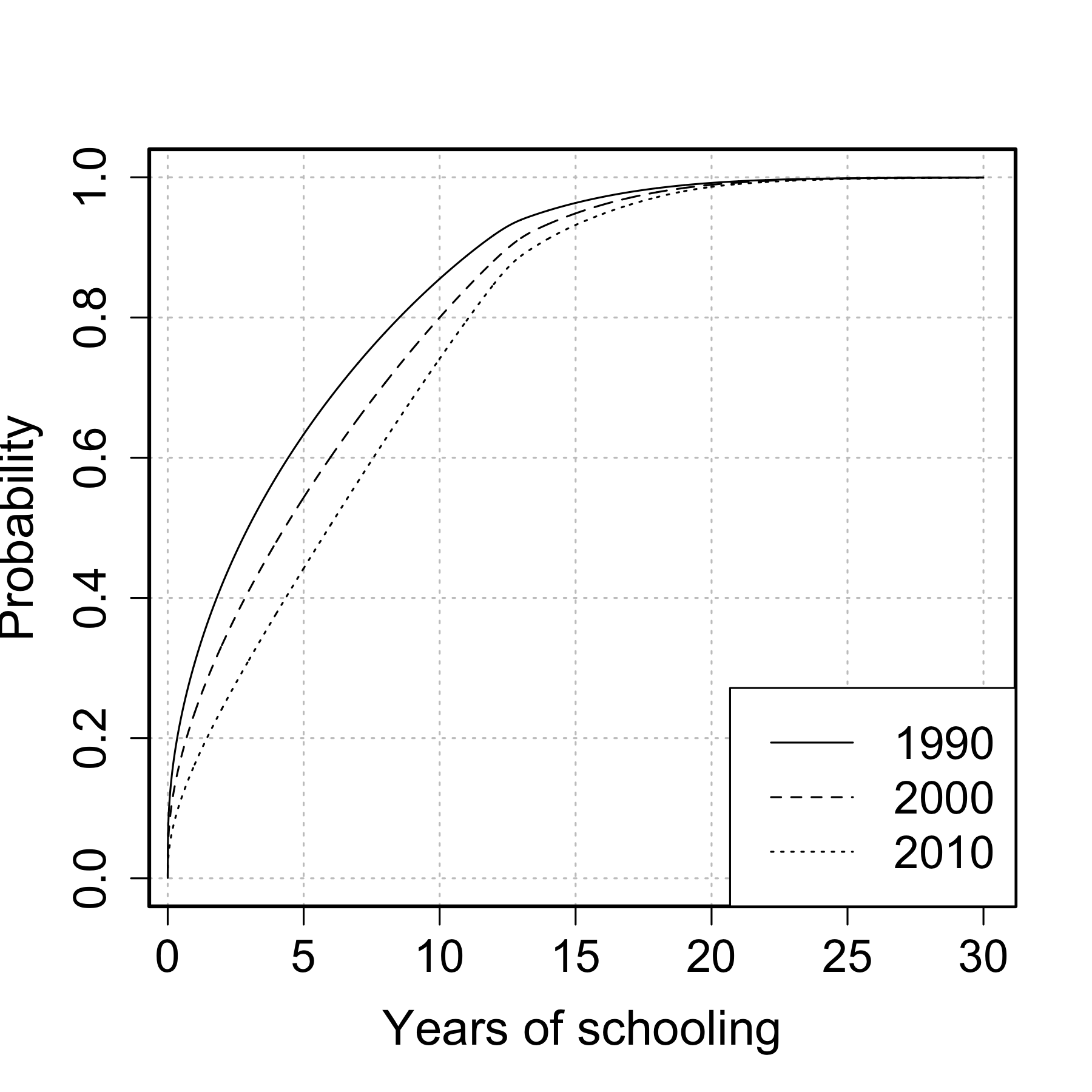}  & \includegraphics[scale = 0.1]{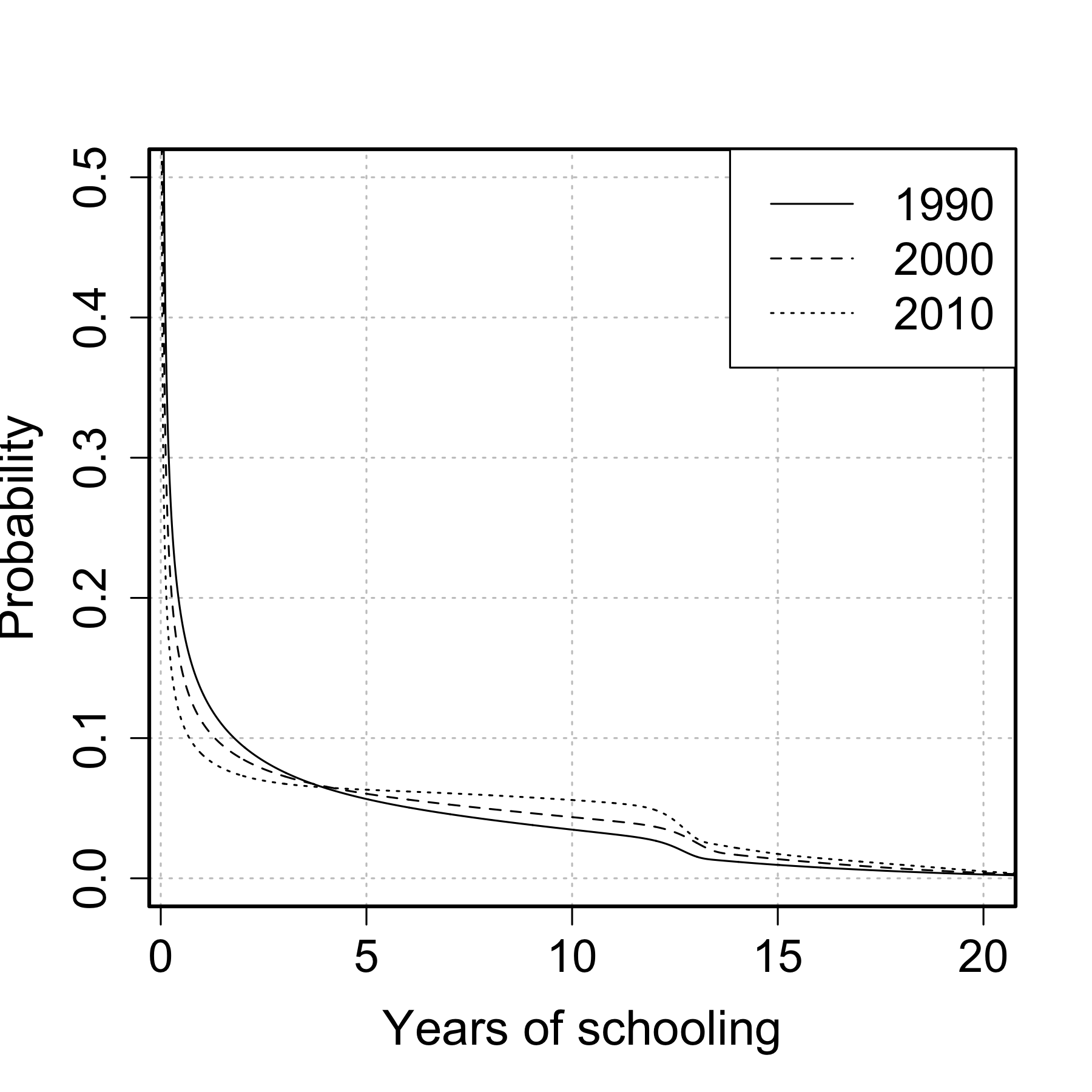}\\
    \includegraphics[scale = 0.1]{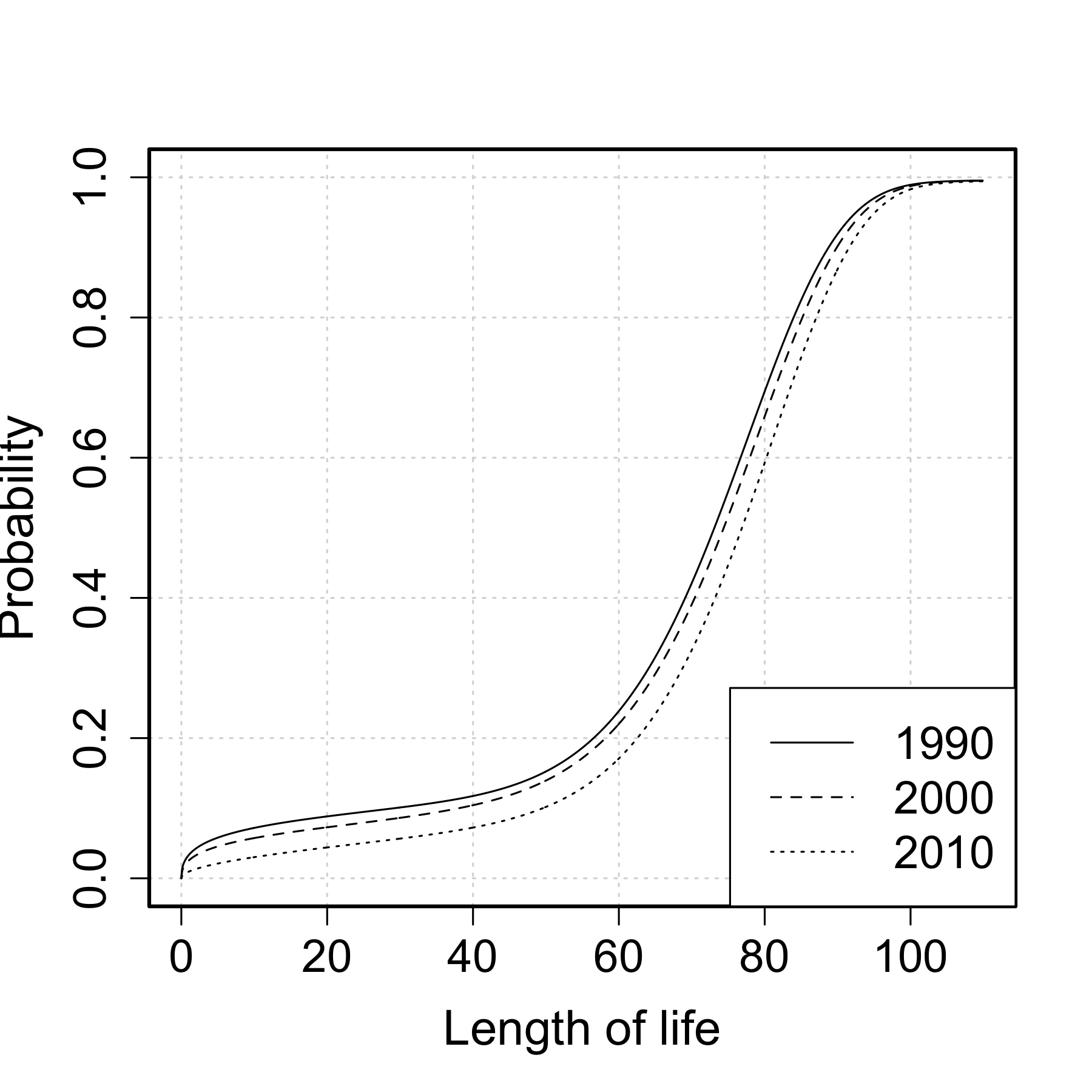}  & \includegraphics[scale = 0.1]{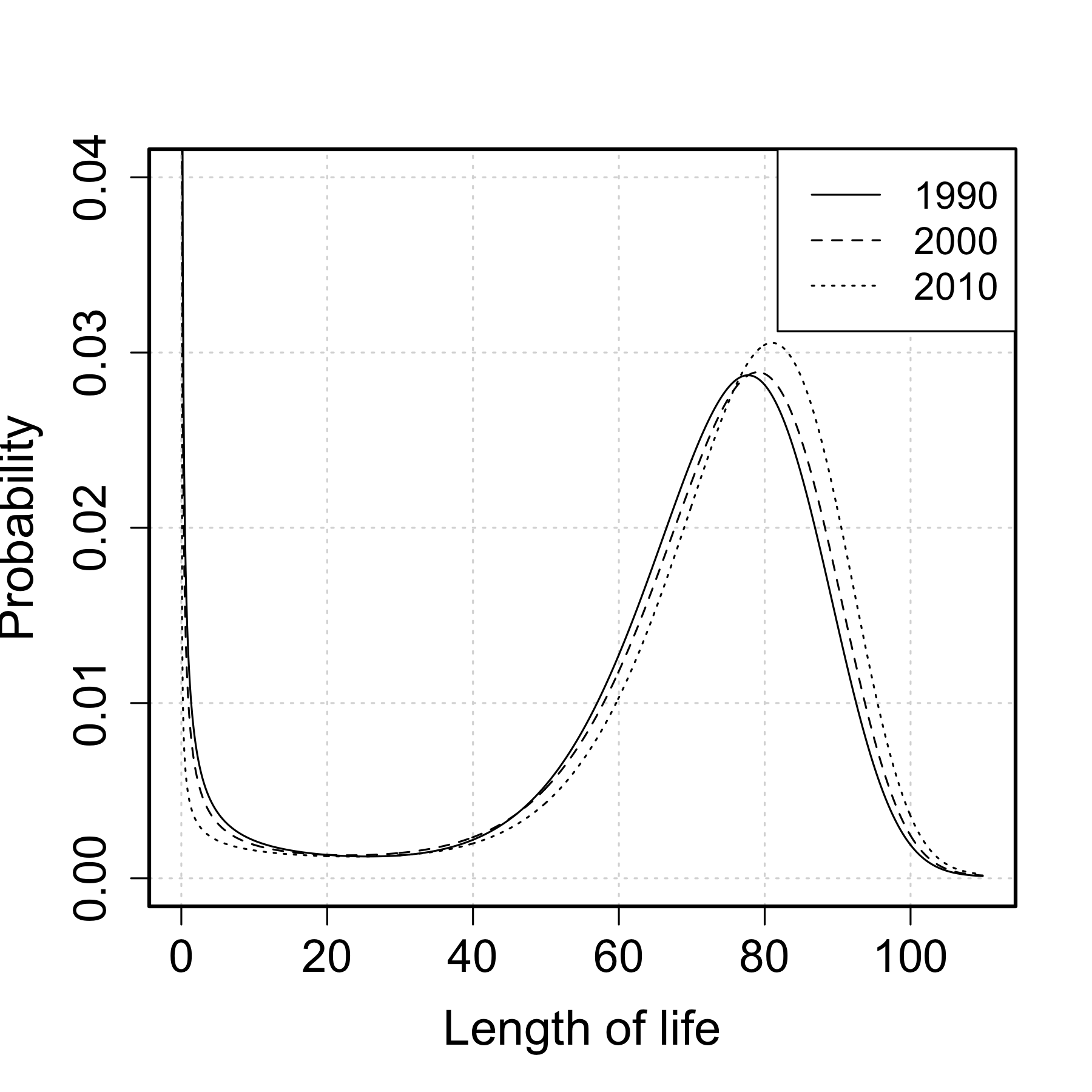}\\
 \end{tabular}
\end{figure}

After discussing the model's goodness-of-fit, we now turn to an analysis of well-being inequality. Although the primary aim of this paper is to examine the evolution of multidimensional inequality in well-being, a dimension-by-dimension analysis also provides valuable insights. Figure \ref{Figure3} displays the global distributions of income, health, and education for the years 1990, 2000, and 2010. Table \ref{ineq_uni} shows the evolution of the unidimensional Atkinson index for $\epsilon = 0.5$ from 1980 to 2015.

In the 1990s, the global income distribution exhibited a two-peaked shape, with a primary mode around US\$1,500 and a secondary peak of wealthier individuals around US\$26,000. Between 1990 and 2010, this distribution gradually transitioned to a bell shape, with a single mode emerging around US\$6,000. This shift can largely be attributed to the substantial economic growth of populous countries like China and India, which resulted in a rightward shift of the global income distribution \citep{lakner2016}. Our cumulative distribution function (cdf) estimates indicate first-order dominance of the 2010 distribution over that of 1990, suggesting that income inequality declined during this period, regardless of the value of the inequality aversion parameter. 

\begin{table}[h]
\begin{center}
\caption{\label{ineq_uni}
Global inequality in income, health and education (1980 - 2015)}
\vspace{0.2cm}
\small{
\begin{tabular}{c c c c c c}
\toprule
	&	Income, log-transformed	&	Income	&	Education	&	Health	\\
	\midrule
1980	&	0.010	&	0.439	&	0.322	&	0.088	\\
1985	&	0.008	&	0.397	&	0.303	&	0.080	\\
1990	&	0.007	&	0.385	&	0.297	&	0.072	\\
1995	&	0.006	&	0.365	&	0.271	&	0.069	\\
2000	&	0.005	&	0.356	&	0.241	&	0.060	\\
2005	&	0.005	&	0.331	&	0.210	&	0.050	\\
2010	&	0.005	&	0.301	&	0.192	&	0.043	\\
2015	&	0.004	&	0.279	&	0.179	&	0.035	\\
\bottomrule
\end{tabular}}
\end{center}
\end{table}

We now turn our attention to the global distribution of educational outcomes. In 1990, there was a high proportion of illiterate individuals, particularly in developing countries in sub-Saharan Africa and Asia \citep{MDG14}.\footnote{The large proportion of the population with no or low levels of education is also partly due to the inclusion of the population aged under 15.} By the early 2000s, the proportion of illiterate individuals had decreased substantially, while primary and secondary attainment rates had increased, reflecting convergence in compulsory years of schooling \citep{murtin11} and expanded educational initiatives in East Asia \citep{Baker1997}. Consequently, the shape of the distribution began to show a modest mode around twelve years of schooling. Our results also suggest that tertiary education attainment rates increased, driven not only by developed countries but also by middle-income countries in Asia and Latin America \citep{jorda2017}.These shifts in the global education distribution led to a significant decline in inequality since 1990. With $\epsilon$ set to 0.5, the Atkinson index decreased from 0.322 in 1980 to 0.179 in 2015, indicating a reduction of 58 percent. 

Turning to the evolution of the global distribution of lifespans, we observe distinct patterns for child and adult populations, shaped by different underlying mortality factors. The evolution of the probability density function (pdf) suggests a substantial reduction in infant mortality rates. Advances in medical science and the expansion of public health policies, such as mass vaccination campaigns, the development of antibiotics and antivirals, improved childbirth practices, and access to clean water and sanitation, are essential in understanding this decline \citep{liu2015}. Additionally, progress in women’s education has been critical in reducing child mortality \citep{cutler2006, gakidou2010}.

Our estimates also show significant improvements in adult lifespan. Progress in this group is closely linked to higher income levels and better education, both of which enhance life expectancy \citep{adams2003}. Therefore, improvements in the other two dimensions of well-being have likely contributed to gains in adult lifespan. Mortality due to HIV/AIDS and malaria has also declined, though at a slower rate, as antiretroviral drugs remain less accessible in many developing countries, and the use of insecticide-treated bednets faces accessibility and usage challenges in malaria-endemic regions \citep{cutler2006}. These advances in reducing the main causes of mortality have contributed to the decline in global inequality in lifespan. In 1980, the Atkinson index was 0.088 for $\epsilon = 0.5$; by 2015, this value fell to 0.035. These figures suggest an overall decline of approximately 60 percent in global inequality in lifespan.

\subsection{Global multidimensional inequality in well-being}

\begin{figure}\caption{Global inequality bands for different values of $\epsilon$ and $\beta$: lower bound (independence copula - blue) and upper bound (comonotonic copula - orange))  \label{Figure4}}
\begin{tabular}{cc}
 \includegraphics[scale = 0.55]{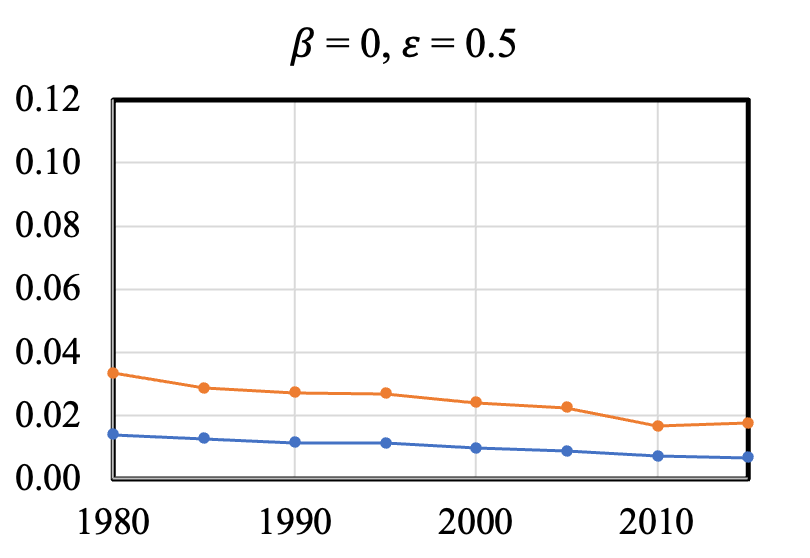}  & \includegraphics[scale = 0.55]{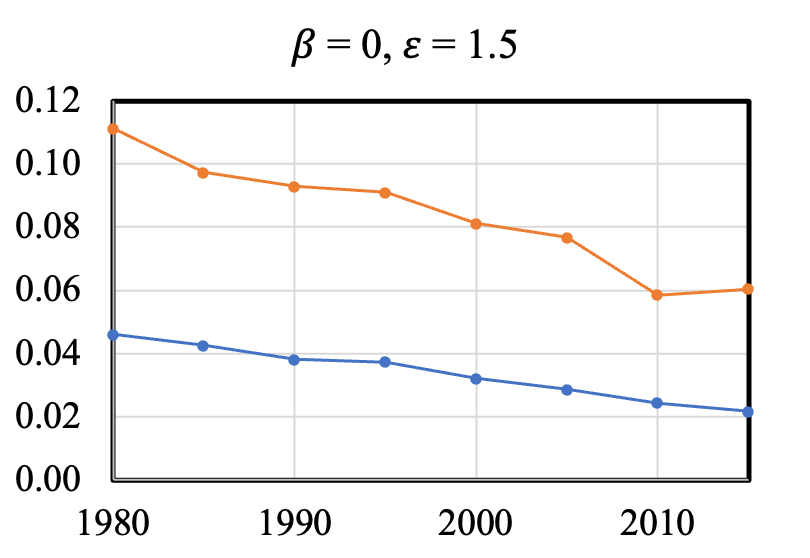}\\
  \includegraphics[scale = 0.55]{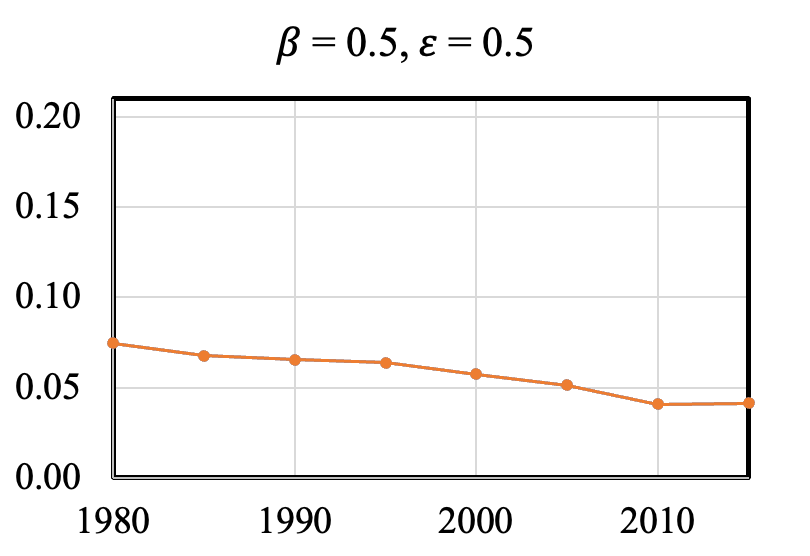}  & \includegraphics[scale = 0.55]{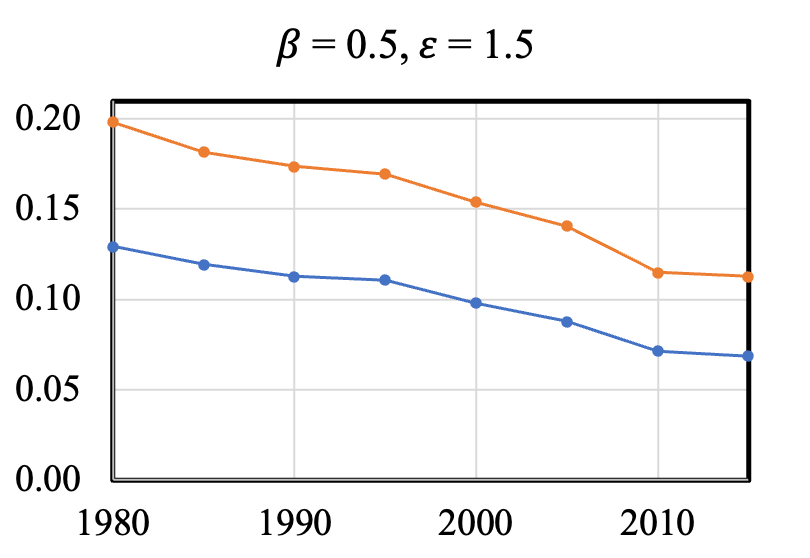}\\
    \includegraphics[scale = 0.55]{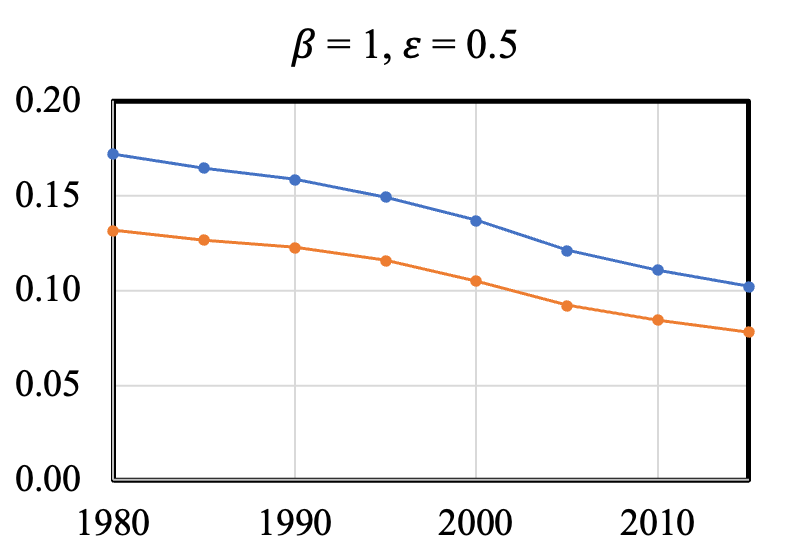}  & \includegraphics[scale = 0.55]{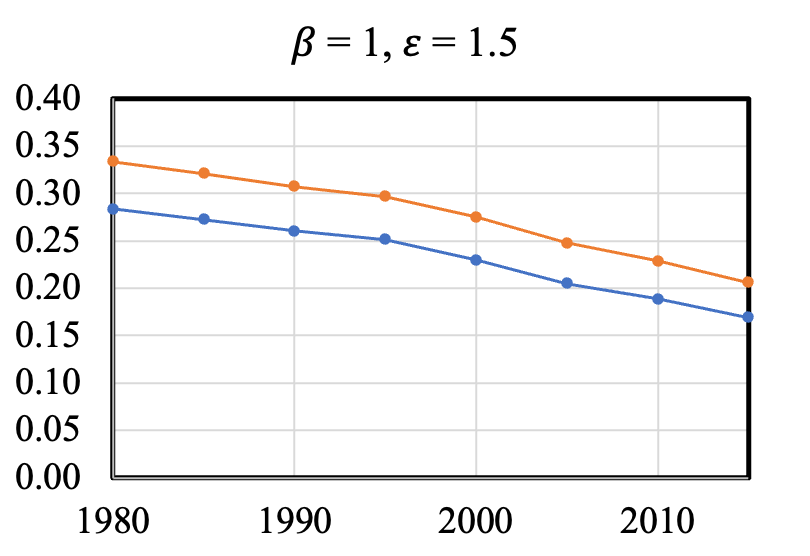}\\
 \end{tabular}
\end{figure}

Following the analysis of unidimensional disparities across income, health, and education, we now turn to examining multidimensional inequality in well-being by considering the combined effects of these dimensions. To capture the complexity of multidimensional inequality, we rely on the Atkinson index, which is sensitive to both inequality aversion (captured by the parameter $\epsilon$) and the degree of complementarity or substitutability between dimensions (captured by the parameter $\beta$). Higher values of $\epsilon$ place more emphasis on the well-being distribution of the least advantaged individuals, while the parameter $\beta$ controls the substitutability between dimensions, with higher values indicating stronger complementarity between the dimensions.

Figure \ref{Figure4} displays global inequality bands for various combinations of $\epsilon$ and $\beta$, with lower bounds represented by the independence copula (blue) and upper bounds by the comonotonic copula (orange). The influence of $\beta$ on the rate of inequality reduction over time is particularly evident in the figure. Specifically, the parameter $\beta$, rather than $\epsilon$, significantly affects the rate of reduction in multidimensional inequality. When $\beta = 0$, we observe an approximate 50 percent reduction in inequality from 1980 to 2015, while for $\beta = 0.5$, the reduction is around 45 percent, and for $\beta = 1$, the reduction falls to about 40 percent. This trend suggests that higher values of $\beta$, which imply greater complementarity between dimensions, lead to a slower decline in inequality over time.

These results also confirm that when $\beta$ exceeds $\epsilon$, the inequality measure behaves inversely to the degree of dependence between dimensions. In this scenario, inequality is higher when dependence is weaker (as modeled by the independence copula) and lower when dependence is stronger (as modeled by the comonotonic copula). This explains why, in cases where $\beta > \epsilon$, the distribution characterized by the comonotonic copula (orange line) shows less inequality than the distribution constructed with the independence copula (blue line).

Furthermore, our estimates suggest that the width of the inequality bands is determined by the difference between $\epsilon$ and $\beta$. When $\epsilon$ and $\beta$ are equal, the Atkinson index becomes insensitive to the dependence structure between dimensions, producing a single line rather than a band in the graph. This occurs because, in this configuration, the multidimensional inequality measure depends only on the marginal distributions and is unaffected by the choice of copula. In cases where $\beta$ is substantially lower than $\epsilon$, the multidimensional inequality measure becomes highly responsive to the assumed dependence structure between dimensions, leading to a wider inequality band. This greater sensitivity arises because, with low $\beta$, the dimensions are treated as near-perfect substitutes, meaning disparities in one dimension can be more easily offset by gains in another. However, the high inequality aversion parameter $\epsilon$ places increased weight on the outcomes of the worse-off individuals, which amplifies the effect of dependence assumptions on the inequality measure. As a result, the combination of low $\beta$ and high $\epsilon$ generates a broader range of possible inequality outcomes, as shown by the widened band between the independence and comonotonic copulas.

When $\beta = 0$ and $\epsilon = 1.5$ or $\epsilon = 0.5$, the inequality bands from 1980 to 2015 are so wide that we cannot reject the hypothesis that inequality remained constant over this period. In this configuration, a nearly straight line could be drawn between the upper and lower bounds of the inequality bands, suggesting there is no clear evidence that multidimensional inequality has either increased or decreased significantly over the decades. This naturally raises the question: how plausible is it that inequality truly remained constant during the analyzed period? In other words, what degree of change in the dependence structure between dimensions would be required to maintain a constant level of global well-being inequality?

To address these questions, we use the copula in Eq. (\ref{mix_c}) to analyze variations in inequality levels for different values of the $\omega$ parameter. As noted earlier, $\omega$ represents the Spearman's correlation index for this copula, meaning that higher values of $\omega$ indicate stronger multivariate dependence between the dimensions. When $\omega = 0$, we have the independent copula, so the inequality level corresponds to the values shown by the blue lines in Figure \ref{Figure4}. Conversely, $\omega = 1$ results in the comonotonic copula, representing perfect positive dependence between dimensions. By varying $\omega$ between these extremes, we can observe how changes in dependence structure impact global well-being inequality and assess the conditions under which inequality might appear stable over time.

\begin{figure}\caption{Global inequality levels for different dependence sturctures \label{omega}}
\begin{center}
 \includegraphics[scale = 0.85]{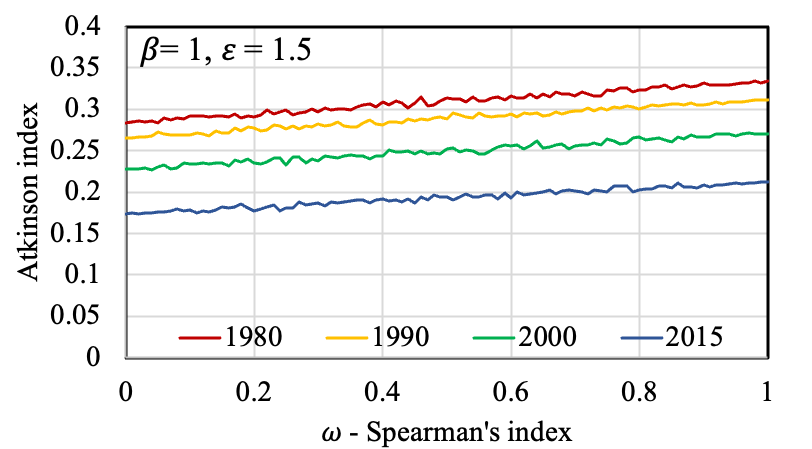}  
  \includegraphics[scale = 0.85]{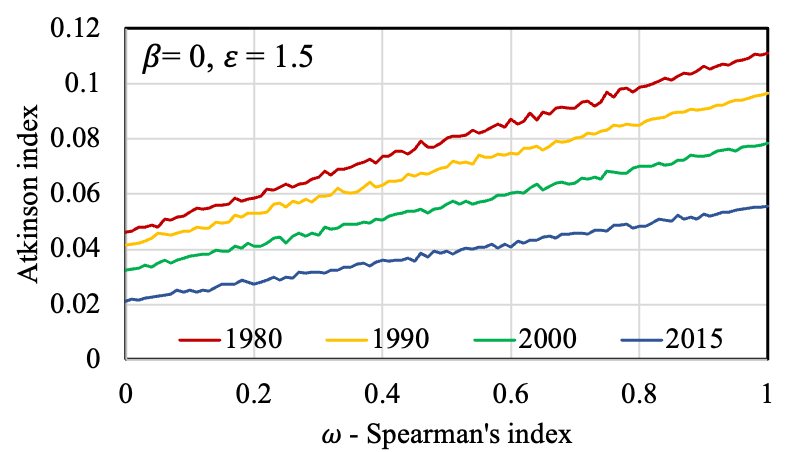} 
\end{center}
\end{figure}

Figure \ref{omega} illustrates the global inequality levels for various dependence structures, with the Atkinson index shown as a function of the $\omega$ parameter, which represents Spearman's correlation. The top panel displays results for $\beta = 1$ and $\epsilon = 1.5$, while the bottom panel shows results for $\beta = 0$ and $\epsilon = 1.5$. Each line corresponds to a specific year --1980, 1990, 2000, and 2015--allowing us to observe how inequality levels change over time under different assumptions about the dependence structure.

In the top panel, where $\beta = 1$, the slope of the lines indicates a gradual reduction in inequality over time, regardless of the value of $\omega$. This suggests that for this level of complementarity (higher $\beta$), inequality decreases consistently, even as dependence increases. Thus, with $\beta = 1$, multidimensional inequality would necessarily fall from 1980 to 2015.

In contrast, the bottom panel, where $\beta = 0$, shows a much more pronounced sensitivity to changes in dependence. Here, if we move from a situation of zero dependence in 1980 (i.e., $\omega = 0$) to a Spearman's index of 0.73, inequality levels would remain constant from 1980 to 2015. Above a Spearman’s index of 0.77, however, inequality would actually increase over this period. This indicates that, when dimensions are perfect substitutes ($\beta = 0$) and the inequality aversion is high, the effect of dependence on inequality is stronger. In such cases, shifts toward greater dependence can offset or even reverse reductions in inequality over time.

\section{Conclusion}

We use a copula-based model to estimate the global multivariate distribution of income, length of life, and schooling from 1980 to 2015. Copulas are a powerful and practical tool for modeling complex multivariate distributions, allowing for different functional forms for each dimension. However, the dependence structure of the multivariate well-being distribution remains unknown, as existing datasets only provide information on the marginal distributions. Therefore, we rely on the independence and comonotonic copulas to compute global inequality bands under these assumptions.

Using a multidimensional Atkinson index, we consider various levels of inequality aversion ($\epsilon$) and complementarity between dimensions ($\beta$), offering insights into how global inequality has evolved under different dependence structures. Our results reveal a general decline in multidimensional inequality over time, largely driven by improvements in health and education outcomes in developing regions and strong economic growth in populous countries such as China and India. However, the rate of inequality reduction is sensitive to the level of complementarity between dimensions. Higher values of $\beta$ indicate a more complementary relationship between dimensions, resulting in slower inequality reduction, while lower values of $\beta$ suggest a more substantial decline as dimensions are treated as near-perfect substitutes.

The study also highlights the importance of the dependence structure. We find that inequality measures are particularly responsive to the level of dependence when $\beta$ is low relative to $\epsilon$, widening the range of possible inequality outcomes. When $\beta$ and $\epsilon$ are equal, the measure becomes invariant to the dependence level, leading to a stable inequality that is determined solely by marginal distributions. This finding has important policy implications: reducing inequality in one dimension, such as income, will not have the same effect on overall inequality if there are strong dependencies across dimensions. As global inequalities in each dimension continue to decline, policymakers may need to consider interactions across dimensions to ensure sustainable and balanced improvements in well-being, especially in regions where dependencies may slow progress.

\section*{Acknowledgements}
 Vanesa Jorda acknowledges financial support from Fundación Ramón Areces (CISP20A6658).

\newpage
\bibliographystyle{ecta}
\bibliography{draft1}

\begin{thebibliography}{58}
\newcommand{\enquote}[1]{``#1''}
\expandafter\ifx\csname natexlab\endcsname\relax\def\natexlab#1{#1}\fi

\bibitem[\protect\citeauthoryear{Adams, Hurd, McFadden, Merrill, and
  Ribeiro}{Adams et~al.}{2003}]{adams2003}
\textsc{Adams, P., M.~D. Hurd, D.~McFadden, A.~Merrill, and T.~Ribeiro} (2003):
  \enquote{Healthy, wealthy, and wise? Tests for direct causal paths between
  health and socioeconomic status,} \emph{Journal of Econometrics}, 112, 3--56.

\bibitem[\protect\citeauthoryear{Alkire}{Alkire}{2002}]{alkire2002}
\textsc{Alkire, S.} (2002): \enquote{Dimensions of human development,}
  \emph{World development}, 30, 181--205.

\bibitem[\protect\citeauthoryear{Alkire and Foster}{Alkire and
  Foster}{2010}]{alkire2010}
\textsc{Alkire, S. and J.~E. Foster} (2010): \enquote{Designing the
  inequality-adjusted human development index,} .

\bibitem[\protect\citeauthoryear{Anand and Segal}{Anand and
  Segal}{2008}]{anand2008}
\textsc{Anand, S. and P.~Segal} (2008): \enquote{{W}hat do we know about global
  income inequality?} \emph{Journal of Economic Literature}, 46, 57--94.

\bibitem[\protect\citeauthoryear{Anand and Segal}{Anand and
  Segal}{2016}]{anand2016}
---\hspace{-.1pt}---\hspace{-.1pt}--- (2016): \enquote{{W}ho are the global top
  1\%?} \emph{International Development Institute Working Paper}.

\bibitem[\protect\citeauthoryear{Atkinson and Brandolini}{Atkinson and
  Brandolini}{2010}]{atkinson2010}
\textsc{Atkinson, A.~B. and A.~Brandolini} (2010): \enquote{`{O}n analyzing the
  world distribution of income',} \emph{The World Bank Economic Review},
  lhp020.

\bibitem[\protect\citeauthoryear{Atkinson et~al.}{Atkinson
  et~al.}{1970}]{atkinson1970}
\textsc{Atkinson, A.~B. et~al.} (1970): \enquote{On the measurement of
  inequality,} \emph{Journal of economic theory}, 2, 244--263.

\bibitem[\protect\citeauthoryear{Baker and Holsinger}{Baker and
  Holsinger}{1997}]{Baker1997}
\textsc{Baker, D.~P. and D.~B. Holsinger} (1997): \enquote{Human capital
  formation and school expansion in Asia: Does a unique regional model exist,}
  \emph{The challenge of Eastern Asian education: Implications for America},
  115--131.

\bibitem[\protect\citeauthoryear{Barro and Lee}{Barro and
  Lee}{2013}]{barro2013}
\textsc{Barro, R.~J. and J.~W. Lee} (2013): \enquote{A new data set of
  educational attainment in the world, 1950--2010,} \emph{Journal of
  development economics}, 104, 184--198.

\bibitem[\protect\citeauthoryear{Becker and Chiswick}{Becker and
  Chiswick}{1966}]{becker1966}
\textsc{Becker, G.~S. and B.~R. Chiswick} (1966): \enquote{Education and the
  Distribution of Earnings,} \emph{The American Economic Review}, 56, 358--369.

\bibitem[\protect\citeauthoryear{Bhalla}{Bhalla}{2002}]{bhalla2002}
\textsc{Bhalla, S.~S.} (2002): \emph{Imagine there's no country: {P}overty,
  inequality, and growth in the era of globalization}, Washington, DC: Peterson
  Institute.

\bibitem[\protect\citeauthoryear{Bosmans, Decancq, and Ooghe}{Bosmans
  et~al.}{2015}]{bosmans2015}
\textsc{Bosmans, K., K.~Decancq, and E.~Ooghe} (2015): \enquote{What do
  normative indices of multidimensional inequality really measure?}
  \emph{Journal of Public Economics}, 130, 94--104.

\bibitem[\protect\citeauthoryear{Bourguignon}{Bourguignon}{1999}]{bourguignon1999}
\textsc{Bourguignon, F.} (1999): \enquote{Comment on `Multidimensioned
  approaches to welfare analysis' by E. Maasoumi,} \emph{Handbook of income
  inequality measurement, Kluwer Academic, London}, 477--84.

\bibitem[\protect\citeauthoryear{Bourguignon and Morrisson}{Bourguignon and
  Morrisson}{2002}]{bourguignon2002}
\textsc{Bourguignon, F. and C.~Morrisson} (2002): \enquote{Inequality among
  world citizens: 1820--1992,} \emph{The American Economic Review}, 92,
  727--744.

\bibitem[\protect\citeauthoryear{Cohen and Soto}{Cohen and
  Soto}{2007}]{cohen07}
\textsc{Cohen, D. and M.~Soto} (2007): \enquote{Growth and human capital: good
  data, good results,} \emph{Journal of Economic Growth}, 12, 51--76.

\bibitem[\protect\citeauthoryear{Cutler, Deaton, and Lleras-Muney}{Cutler
  et~al.}{2006}]{cutler2006}
\textsc{Cutler, D., A.~Deaton, and A.~Lleras-Muney} (2006): \enquote{The
  determinants of mortality,} \emph{The Journal of Economic Perspectives}, 20,
  97--120.

\bibitem[\protect\citeauthoryear{Decancq}{Decancq}{2011}]{decancq2011}
\textsc{Decancq, K.} (2011): \enquote{Measuring global well-being inequality: A
  dimension-by-dimension or multidimensional approach?} \emph{Reflets et
  perspectives de la vie {\'e}conomique}, 50, 179--196.

\bibitem[\protect\citeauthoryear{Decancq}{Decancq}{2014}]{decancq2014}
---\hspace{-.1pt}---\hspace{-.1pt}--- (2014): \enquote{Copula-based measurement
  of dependence between dimensions of well-being,} \emph{Oxford Economic
  Papers}, 66, 681--701.

\bibitem[\protect\citeauthoryear{Decancq and Lugo}{Decancq and
  Lugo}{2012}]{decancq2013}
\textsc{Decancq, K. and M.~A. Lugo} (2012): \enquote{Inequality of wellbeing: A
  multidimensional approach,} \emph{Economica}, 79, 721--746.

\bibitem[\protect\citeauthoryear{Decancq and Schokkaert}{Decancq and
  Schokkaert}{2016}]{decancq2017}
\textsc{Decancq, K. and E.~Schokkaert} (2016): \enquote{Beyond GDP: Using
  equivalent incomes to measure well-being in Europe,} \emph{Social indicators
  research}, 126, 21--55.

\bibitem[\protect\citeauthoryear{DESA}{DESA}{2017}]{desa2017}
\textsc{DESA, U.} (2017): \enquote{World population prospects: the 2017
  revision,} \emph{Population division of the department of economic and social
  affairs of the United Nations Secretariat, New York}.

\bibitem[\protect\citeauthoryear{Dhongde and Minoiu}{Dhongde and
  Minoiu}{2013}]{dhongde2013}
\textsc{Dhongde, S. and C.~Minoiu} (2013): \enquote{Global poverty estimates: A
  sensitivity analysis,} \emph{World Development}, 44, 1--13.

\bibitem[\protect\citeauthoryear{Dowrick and Akmal}{Dowrick and
  Akmal}{2005}]{dowrick2005}
\textsc{Dowrick, S. and M.~Akmal} (2005): \enquote{`{C}ontradictory trends in
  global income inequality: {A} tale of two biases',} \emph{Review of Income
  and Wealth}, 51, 201--229.

\bibitem[\protect\citeauthoryear{Foster, Lopez-Calva, and Szekely}{Foster
  et~al.}{2005}]{foster2005}
\textsc{Foster, J.~E., L.~F. Lopez-Calva, and M.~Szekely} (2005):
  \enquote{Measuring the distribution of human development: methodology and an
  application to Mexico,} \emph{Journal of Human Development}, 6, 5--25.

\bibitem[\protect\citeauthoryear{Gakidou, Cowling, Lozano, and Murray}{Gakidou
  et~al.}{2010}]{gakidou2010}
\textsc{Gakidou, E., K.~Cowling, R.~Lozano, and C.~J. Murray} (2010):
  \enquote{Increased educational attainment and its effect on child mortality
  in 175 countries between 1970 and 2009: a systematic analysis,} \emph{The
  Lancet}, 376, 959--974.

\bibitem[\protect\citeauthoryear{Grimm, Harttgen, Klasen, and Misselhorn}{Grimm
  et~al.}{2008}]{grimm2008}
\textsc{Grimm, M., K.~Harttgen, S.~Klasen, and M.~Misselhorn} (2008):
  \enquote{A human development index by income groups,} \emph{World
  Development}, 36, 2527--2546.

\bibitem[\protect\citeauthoryear{Harttgen and Klasen}{Harttgen and
  Klasen}{2012}]{harttgen2012}
\textsc{Harttgen, K. and S.~Klasen} (2012): \enquote{A household-based human
  development index,} \emph{World Development}, 40, 878--899.

\bibitem[\protect\citeauthoryear{Jenkins}{Jenkins}{2009}]{jenkins09}
\textsc{Jenkins, S.~P.} (2009): \enquote{Distributionally-sensitive inequality
  indices and the GB2 income distribution,} \emph{Review of Income and Wealth},
  55, 392--398.

\bibitem[\protect\citeauthoryear{Jord{\'a} and Alonso}{Jord{\'a} and
  Alonso}{2017}]{jorda2017}
\textsc{Jord{\'a}, V. and J.~M. Alonso} (2017): \enquote{New estimates on
  educational attainment using a continuous approach (1970--2010),} \emph{World
  Development}, 90, 281--293.

\bibitem[\protect\citeauthoryear{Jorda and Ni{\~n}o-Zaraz{\'u}a}{Jorda and
  Ni{\~n}o-Zaraz{\'u}a}{2019}]{jorda2019}
\textsc{Jorda, V. and M.~Ni{\~n}o-Zaraz{\'u}a} (2019): \enquote{Global
  inequality: How large is the effect of top incomes?} \emph{World
  Development}, 123, 104593.

\bibitem[\protect\citeauthoryear{Jorda, Sarabia, and J{\"a}ntti}{Jorda
  et~al.}{2018}]{jorda2020}
\textsc{Jorda, V., J.~M. Sarabia, and M.~J{\"a}ntti} (2018):
  \enquote{Estimation of income inequality from grouped data,} \emph{arXiv
  preprint arXiv:1808.09831}.

\bibitem[\protect\citeauthoryear{Kleiber and Kotz}{Kleiber and
  Kotz}{2003}]{kleiber2003}
\textsc{Kleiber, C. and S.~Kotz} (2003): \emph{Statistical size distributions
  in economics and actuarial sciences}, vol. 470, New Jersey: John Wiley \&
  Sons.

\bibitem[\protect\citeauthoryear{Kummu, Taka, and Guillaume}{Kummu
  et~al.}{2018}]{kummu2018}
\textsc{Kummu, M., M.~Taka, and J.~H. Guillaume} (2018): \enquote{Gridded
  global datasets for gross domestic product and Human Development Index over
  1990--2015,} \emph{Scientific data}, 5, 1--15.

\bibitem[\protect\citeauthoryear{Lakner and Milanovic}{Lakner and
  Milanovic}{2016}]{lakner2016}
\textsc{Lakner, C. and B.~Milanovic} (2016): \enquote{Global income
  distribution from the fall of the Berlin Wall to the Great Recession,}
  \emph{World Bank Economic Review}, 30, 203--232.

\bibitem[\protect\citeauthoryear{Liu, Oza, Hogan, Perin, Rudan, Lawn, Cousens,
  Mathers, and Black}{Liu et~al.}{2015}]{liu2015}
\textsc{Liu, L., S.~Oza, D.~Hogan, J.~Perin, I.~Rudan, J.~E. Lawn, S.~Cousens,
  C.~Mathers, and R.~E. Black} (2015): \enquote{Global, regional, and national
  causes of child mortality in 2000--13, with projections to inform post-2015
  priorities: an updated systematic analysis,} \emph{The Lancet}, 385,
  430--440.

\bibitem[\protect\citeauthoryear{McDonald}{McDonald}{1984}]{Mcdonald1984}
\textsc{McDonald, J.~B.} (1984): \enquote{Some Generalized Functions for the
  Size Distribution of Income,} \emph{Econometrica}, 52, 647--665.

\bibitem[\protect\citeauthoryear{McDonald and Mantrala}{McDonald and
  Mantrala}{1995}]{mcdonald1995m}
\textsc{McDonald, J.~B. and A.~Mantrala} (1995): \enquote{The distribution of
  personal income: revisited,} \emph{Journal of Applied Econometrics}, 10,
  201--204.

\bibitem[\protect\citeauthoryear{McDonald and Xu}{McDonald and
  Xu}{1995}]{mcdonald1995}
\textsc{McDonald, J.~B. and Y.~J. Xu} (1995): \enquote{A generalization of the
  beta distribution with applications,} \emph{Journal of Econometrics}, 66,
  133--152.

\bibitem[\protect\citeauthoryear{McGillivray and Markova}{McGillivray and
  Markova}{2010}]{mcgillivray2010}
\textsc{McGillivray, M. and N.~Markova} (2010): \enquote{Global inequality in
  well-being dimensions,} \emph{The Journal of Development Studies}, 46,
  371--378.

\bibitem[\protect\citeauthoryear{McGillivray and Pillarisetti}{McGillivray and
  Pillarisetti}{2004}]{mcgillivray2004}
\textsc{McGillivray, M. and J.~R. Pillarisetti} (2004): \enquote{International
  inequality in well-being,} \emph{Journal of International Development}, 16,
  563--574.

\bibitem[\protect\citeauthoryear{Milanovic}{Milanovic}{2011}]{milanovic2011}
\textsc{Milanovic, B.} (2011): \emph{Worlds apart: {M}easuring international
  and global inequality}, Princeton University Press.

\bibitem[\protect\citeauthoryear{Miller}{Miller}{1960}]{miller1960}
\textsc{Miller, H.~P.} (1960): \enquote{Annual and lifetime income in relation
  to education: 1939-1959,} \emph{The American Economic Review}, 50, 962--986.

\bibitem[\protect\citeauthoryear{Morrisson and Murtin}{Morrisson and
  Murtin}{2009}]{morrisson09}
\textsc{Morrisson, C. and F.~Murtin} (2009): \enquote{The Century of
  Education,} \emph{Journal of Human Capital}, 3, 1--42.

\bibitem[\protect\citeauthoryear{Murtin and Viarengo}{Murtin and
  Viarengo}{2011}]{murtin11}
\textsc{Murtin, F. and M.~Viarengo} (2011): \enquote{The expansion and
  convergence of compulsory schooling in Western Europe, 1950--2000,}
  \emph{Economica}, 78, 501--522.

\bibitem[\protect\citeauthoryear{Ni{\~n}o-Zaraz{\'u}a, Roope, and
  Tarp}{Ni{\~n}o-Zaraz{\'u}a et~al.}{2017}]{nino2014}
\textsc{Ni{\~n}o-Zaraz{\'u}a, M., L.~Roope, and F.~Tarp} (2017):
  \enquote{Global inequality: {Relatively} lower, absolutely higher,}
  \emph{Review of Income and Wealth}, 63, 661--684.

\bibitem[\protect\citeauthoryear{P{\'e}rez and Prieto-Alaiz}{P{\'e}rez and
  Prieto-Alaiz}{2016}]{perez2016}
\textsc{P{\'e}rez, A. and M.~Prieto-Alaiz} (2016): \enquote{Measuring the
  dependence among dimensions of welfare: A study based on Spearman's footrule
  and Gini's gamma,} \emph{International Journal of Uncertainty, Fuzziness and
  Knowledge-Based Systems}, 24, 87--105.

\bibitem[\protect\citeauthoryear{Ram}{Ram}{1992}]{ram1992}
\textsc{Ram, R.} (1992): \enquote{Intercountry inequalities in income and
  basic-needs indicators: A recent perspective,} \emph{World Development}, 20,
  899--905.

\bibitem[\protect\citeauthoryear{Shorrocks and Wan}{Shorrocks and
  Wan}{2008}]{Shorrocks08}
\textsc{Shorrocks, A. and G.~Wan} (2008): \enquote{{Ungrouping Income
  Distributions: Synthesising Samples for Inequality and Poverty Analysis},}
  Working Paper Series RP2008/16, World Institute for Development Economic
  Research (UNU-WIDER).

\bibitem[\protect\citeauthoryear{Sklar}{Sklar}{1959}]{sklar1959}
\textsc{Sklar, M.} (1959): \enquote{Fonctions de repartition an dimensions et
  leurs marges,} \emph{Publ. inst. statist. univ. Paris}, 8, 229--231.

\bibitem[\protect\citeauthoryear{Smits and Monden}{Smits and
  Monden}{2009}]{smits2009}
\textsc{Smits, J. and C.~Monden} (2009): \enquote{Length of life inequality
  around the globe,} \emph{Social Science \& Medicine}, 68, 1114--1123.

\bibitem[\protect\citeauthoryear{Smits and Permanyer}{Smits and
  Permanyer}{2019}]{smits2019}
\textsc{Smits, J. and I.~Permanyer} (2019): \enquote{The subnational human
  development database,} \emph{Scientific data}, 6, 1--15.

\bibitem[\protect\citeauthoryear{Stacy}{Stacy}{1962}]{Stacy1962}
\textsc{Stacy, E.~W.} (1962): \enquote{A generalization of the gamma
  distribution,} \emph{The Annals of Mathematical Statistics}, 1187--1192.

\bibitem[\protect\citeauthoryear{Stiglitz, Sen, and Fitoussi}{Stiglitz
  et~al.}{2010}]{stiglitz2009}
\textsc{Stiglitz, J., A.~Sen, and J.~Fitoussi} (2010): \emph{Mismeasuring Our
  Lives: Why GDP Doesn't Add Up}, New York: New Press.

\bibitem[\protect\citeauthoryear{Tkach and Gigliarano}{Tkach and
  Gigliarano}{2018}]{tkach2018}
\textsc{Tkach, K. and C.~Gigliarano} (2018): \enquote{Multidimensional poverty
  measurement: dependence between well-being dimensions using copula function,}
  \emph{Rivista Italiana di Economia Demografia e Statistica}, 72.

\bibitem[\protect\citeauthoryear{UNESCO}{UNESCO}{2013}]{UNESCO2013}
\textsc{UNESCO} (2013): \enquote{UIS methodology for estimation of mean years
  of schooling,} Available at:
  http://www.uis.unesco.org/Education/Documents/mean-years-schooling-indicator-methodology-en.pdf.

\bibitem[\protect\citeauthoryear{Unesco}{Unesco}{2014}]{UNESCO}
\textsc{Unesco} (2014): \enquote{Statistical Yearbook,} \emph{Paris}.

\bibitem[\protect\citeauthoryear{UnitedNations}{UnitedNations}{2014}]{MDG14}
\textsc{UnitedNations} (2014): \emph{The Millennium Development Goals Report
  2008}, United Nations Publications.

\bibitem[\protect\citeauthoryear{UNU-WIDER}{UNU-WIDER}{2014}]{WIID2014}
\textsc{UNU-WIDER} (2014): \enquote{World Income Inequality Database
  (WIID3.0A),} .

\end{thebibliography}

\newpage
\end{document}